\documentstyle[12pt]{article}

\newcommand{\setdoublespace}%
   {\small\normalsize}

\makeatletter 
\def\ps@titlepage{\let\@mkboth\@gobbletwo%
 \def\@oddhead{\hfil%
 \begin{tabular}{r}
   \sc Institut f\"ur Theoretische Kernphysik\\\sc TK 92 08
 \end{tabular}}%
 \def\@oddfoot{\hfil\rm\thepage\hfil}%
 \let\@evenhead\@oddhead\let\@evenfoot\@oddfoot%
 \def\sectionmark##1{}\def\subsectionmark##1{}}
\makeatother

\ifx\mathrm\undefined
  \newcommand{\mathrm}[1]{\mathord{\rm {#1}}}
\fi
\newcommand{\unit}[1]{\mathop{\mathrm{#1}}\nolimits}
\newcommand{\eV}{\unit{eV}}
\newcommand{\keV}{\unit{keV}}
\newcommand{\MeV}{\unit{MeV}}
\newcommand{\GeV}{\unit{GeV}}
\newcommand{\TeV}{\unit{TeV}}
\newcommand{\cm}{\unit{cm}}
\newcommand{\g}{\mathop{\smash{\mathrm{g}}}\nolimits}

\newcommand{\dd}{{\mathrm{d}}}

\newcommand{\Emin}{E^{(\mathrm {min})}}
\newcommand{\Emax}{E^{(\mathrm {max})}}
\newcommand{\Etot}{E^{(\mathrm {tot})}}

\newcommand{\ie}{{\it i.e.}}

\newcommand{\etc}{{\it etc.}}

\begin{document}

\title{Neutrino production through hadronic cascades in AGN accretion disks}
\author{%
Lukas Nellen\\
Institut f\"ur Theoretische Kernphysik\\
Nu{\ss}allee 14--16\\
D-5300 Bonn 1\\
Germany
\and
Karl Mannheim and Peter L. Biermann\\
Max-Planck-Institut f\"ur Radioastronomie\\
Auf dem H\"ugel 69\\
D-5300 Bonn 1\\
Germany}

\maketitle
\thispagestyle{titlepage}

\setdoublespace

\begin{abstract}
We consider the production of neutrinos in active galactic nuclei
(AGN) through hadronic cascades. The initial, high energy nucleons
are accelerated in a source above the accretion disk around the
central black hole.
{}From the source, the particles diffuse back to the disk and initiate
hadronic cascades.
The observable output from the cascade are electromagnetic radiation
and neutrinos.
We use the observed diffuse background X-ray
luminosity, which presumably results from this process, to predict the
diffuse neutrino flux
close to existing limits from the Frejus experiment.
The resulting neutrino spectrum is $E^{-2}$ down to the $\GeV$ region.
We discuss modifications of this scenario which reduce the predicted
neutrino flux.
\end{abstract}

\section{Introduction}

The active nuclei of galaxies (AGN), ranging in luminosity
{}from Seyfert galaxies
to quasars, are the most powerful individual sources of radiation in
the Universe. To explain the power emitted by such objects, one
generally assumes the existence of a central engine in which the
gravitational energy of matter falling into a supermassive black hole
gets converted into radiation.  Even though so far only
electromagnetic radiation has been observed, it is generally assumed
that other particles are accelerated and emitted as well to explain
the tight relationship between the non-thermal and thermal components
in the~UV and~\mbox{X-ray} spectra%
{}~\cite{lw:apj335,pnsgf:nature344,fgmr:mnras242,cla+a:apj393,ckb:aa219}.
Of special interest are neutrinos, since they can travel cosmological
distances without losing the information on the direction they
originated from.  Large underwater detectors%
{}~\cite{spi:phe91-17,ste:hdc-13-91} or
detectors in the antarctic ice cap~\cite{bar+a:mad-ph-634} are
used as neutrino telescopes. Recent calculations have shown that the
flux of neutrinos
originating in an AGN could be detected by such experiments%
{}~\cite{sdss:hawaii1992,sdss:prl66,%
sp:hawaii1992,man:hawaii1992,bie:hawaii1992}.
Data from proton decay experiments~\cite{mey:frejus-limit}
and airshower arrays~\cite{hz:pl289b} is already sensitive enough to
constrain such models significantly.

\section{The AGN model}

In the following we are interested in the production of neutrinos in
radio-quiet AGN\@. In the ``standard'' model for AGN one assumes
that the central black hole is surrounded by an accretion disk of
infalling matter. Besides that, one expects to find bipolar outflow of gas
and plasma perpendicular to the disk (jet). Jets can be seen in
different objects with disk accretion and in many
AGN~\cite{bp:araa22,mr:jiers1991}. One expects shocks in the plasma
of the jets~\cite{man:hawaii1992} which could accelerate protons
through first order Fermi acceleration to energies up to
$O(10^9\GeV)$ with a powerlaw spectrum of $E^{-2}$~\cite{bs:apj322}.
The observation of $\gamma$-rays with the same kind of
spectrum~\cite{sjs:apj390} indicates that indeed there is
shock-acceleration outside the core of the AGN, so that the photons
can escape.

In~\cite{nie:diplom1991}, Niemeyer showed that the far infra-red (FIR)
emission of AGN can be explained by assuming that the accelerated protons
diffuse back from the jet to the accretion disk and heat dust clouds beyond
the outer region of the accretion disk. In the inner part of the disk,
protons hitting the disk will initiate hadronic cascades through
interactions with the gas in the disk. This could feed into an
electromagnetic cascade and thereby generate the observed X-ray and gamma
emission~\cite{bie:hawaii1992}.

To describe the accretion disk, we use the model by Shakura and
Sunyaev~\cite{ss:aa24}.
For our purposes, we will concentrate on the inner region, which is
the radiation dominated part of the disk (neglecting the thin, innermost
ring around the black hole).

\section{Hadronic cascades}

Due to the observed short timescales of the variability of the X-ray and
UV~components of the AGN spectrum, this part of the electro-magnetic
radiation must predominantly originate in a small, central region
around the central black
hole~\cite{sdss:prl66,sdss:hawaii1992} with a typical mass of
$10^8 M_\odot$ for luminous AGN\@.
We expect neutrino production to take place in the same region,
so we need to know the incoming particle flux in this
region. From~\cite{nie:diplom1991} we know that the FIR spectra are best
fitted by assuming that the protons originate from a point source in
the jet at
$z_0\le O(10^{18}\cm)\approx 3\cdot10^4R_S$ above the disk.
We consider two possibilities how protons from this or a similar
source can travel back to the central region of the disk and
initiate hadronic cascades in the accreting gas.

I) The region interesting for the production of neutrinos is much closer
to the centre of the disk; it extends
to about $200 R_S \approx 10^{16}\cm$. In the disk
model~\cite{ss:aa24}, this corresponds to the innermost region of the
disk. An isotropic proton source in the jet, similar to the kind of
source mentioned above, located close to the origin, \ie, at
$z_0\ll 200R_S$, could then deposit at least half its
luminosity into the disk.
This happens simply for geometrical reasons, since the inner disk
occupies about half of the horizon of the proton source.  In this
approximation, there is no dependence on the exact distribution of the
proton source along the jet or on details of the proton transport.

II) It is also possible that the proton source at~$z_0=O(10^{18}\cm)$,
needed for the explanation of the FIR radiation, emits a strong
downward proton flux.
Such an anisotropy in the proton emission arises naturally, since
the interaction of the protons in the source
with upward UV~to X-ray photons from the accretion disk preferably produces
downward neutrons through the reaction
$p\gamma\to\Delta\to n\pi^+$, if the photon scattering in the ambient
medium is negligible.
The upcoming photon density is sufficiently low that
$\tau_{n\gamma}<1$, \ie, the neutrons will reach the disk without
further reactions with photons.
Furthermore, above $50\TeV$, the neutrons will reach the disk without
decaying back into protons.
The incident nucleons will
then induce a hadronic cascade in the inner accretion disk.
In this scenario, we get beaming for high energy neutrinos.
At lower energies, where protons are part of the cascade, the magnetic
field in the disk randomises the orientation of the cascade.
Also, the spectrum of the nucleons emitted from the source is flatter
than~$E^{-2}$ at high energies which leads to a flatter high energy
neutrino spectrum as well.

In the following, we will concentrate on scenario I with a
pure~$E^{-2}$ spectrum for the proton source.
In the inner region of the accretion disk, the column density of the disk is
\begin{equation}
u_0 \approx 4.6\,\alpha^{-1}\left(\dot M \over \dot M_{\rm edd}\right)^{-1}
                 \left(R \over 3R_S\right)^{3/2} {\g \over\cm^2}.
\end{equation}
If we take $\dot M/M_{\rm edd} \approx 0.1$ and
$\alpha \approx 0.1$, we see that the column density is high enough
compared to the mean free path for $pp$-collisions of
$O(50 \g/\cm^2)$ for a hadronic cascade to develop.
The magnetic field confines the protons to the disk, which leads to an
even further increase of the effective column density seen by the
protons. Even though neutrons which are produced in the cascade are
not confined by the magnetic field, the amount of gas present is
sufficient to prevent them from escaping.

The hadronic cascade in the disk is much simpler than cascades in the
earth's atmosphere~\cite{gai:crpp,gy:arnps30,gptm:rmp50},
since the density of the accretion disk is so low that
all unstable particles decay rather than interact. Therefore the cascade
consists of a nucleonic part which feeds into the mesonic and
electro-magnetic channels; no pion-nucleon reactions occur.
Furthermore, the pion channel will always be the dominant channel for
the production of neutrinos. This is different from the production of
cosmic ray neutrinos in the atmosphere, since there reactions of pions
with nuclei remove the pions from the parent population for neutrino
production. This leads to a steeper neutrino spectrum and
to the increasing importance of kaons and
charmed mesons for the neutrino production in the atmosphere
while heavy mesons are of little importance for AGN neutrino
production.
\begin{figure}[tbp]
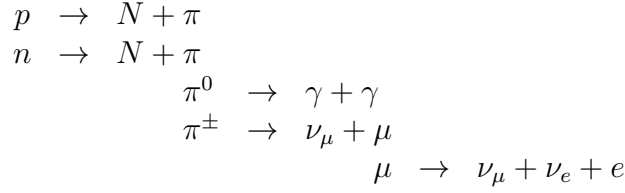

$$
\begin{array}{ccr@{}lcrcl}
p&\to&N+{}&\pi&&&&\\
n&\to&N+{}&\pi&&&&\\
&&&{\pi^0}&\to&\multicolumn{1}{l}{\gamma+\gamma}&&\\
&&&{\pi^\pm}&\to&\nu_\mu + \mu&&\\
&&&&&\mu&\to&\nu_\mu + \nu_e + e
\end{array}
$$
\caption{The structure of the hadronic cascade: $pp$~and
$pn$~interactions drive the cascade and feed the production of pions,
leptons and hard photons.
We do not distinguish particles and anti-particles in this figure. So
the leptonic final channels need to be doubled.}
\label{fig:cascade}
\end{figure}
Under
the assumptions of scaling hadronic cross-sections, the power law of the
incident protons is kept and inherited by all the secondaries, \ie, the
flux of particle $x$ is
\begin{equation}
\dot n_x(E) = C_x\,\dot n_p(E),\qquad \dot n_p(E)=\tilde C\,E^{-\gamma-1},
\end{equation}
where $\tilde C$ is the normalisation of the proton spectrum.
The specific power in  protons is then
\begin{equation}\begin{array}{r@{}c@{}l}
\dot\Etot_p
    &{}={}& \displaystyle \int_{\Emin_p}^{\Emax_p} E' \dot n_p(E') \, \dd E',\\
    &{}={}& \left\{
	\begin{array}{l@{\qquad\mbox{for }}l}
        \tilde C \left(\gamma -1\right)^{-1}
        \left( \Emin_p \over \Emax_p \right)^{-\gamma+1},&
        \gamma \not=1,\\
        \tilde C \ln\left( \Emax_p \over \Emin_p \right), & \gamma=1.\\
	\end{array}\right.
\end{array}\end{equation}
For the secondaries, the energy range is shifted down by a
factor~$\epsilon_x$; the
resulting power is
\begin{equation}\begin{array}{r@{}c@{}l}
\dot\Etot_x
    &{}={}& \displaystyle \int_{\epsilon_x\Emin_p}^{\epsilon_x\Emax_p}
                              E' \dot n_p(E') \, \dd E',\\
    &{}={}& C_x\,\epsilon_x^{-\gamma+1} \dot \Etot_p.
\end{array}
\label{eqn:sec-power}
\end{equation}
\begin{table}
$$
\begin{array}{c|c|c}
$Species$&\epsilon_x&C_x=\Etot_x / \Etot_p\\
\hline\hline
\gamma
	& 0.070 &0.29\\
e
	& 0.056 &0.16\\
\hline
\nu_\mu \mbox{ from $\pi$}
	& 0.040 &0.13\\
\nu_\mu \mbox{ from $\mu$}
	& 0.056 &0.16\\
\nu_e   & 0.048 &0.14
\end{array}
$$
\caption{Relative normalization of the spectrum of secondaries,
shift of particle energies,
and fraction of the incident proton luminosity energy carried by
the stable secondaries for $\gamma=1$.
The total energy in one species is split evenly between particles and
anti-particles.
}
\label{tab:gamma=1}
\end{table}

The cascade equations can be solved approximately to determine~$C_x$
and~$\epsilon_x$.
For scaling hadronic cross-sections, the experimental data is
summarised by the spectrum weighted moments
\begin{equation}
Z_{ab}=\int_0^1 x_L^{\gamma-1} F_{ab}(x_L) \,\dd x_L,
\end{equation}
where $x_L=E_b/E_a$ is the fraction of the energy of the primary
particle transferred to the secondary
and $F_{ab}(x_L)=E_b\,\dd n_b(x_L)/\dd E_b$ is the
dimensionless, inclusive cross-section for the production of particles
of type~$b$. For~$\gamma > 1$, the weighting factor~$x_L^{\gamma-1}$
reduces the importance of the small-$x_F$ region, where the accuracy of
experimental data is low. Unfortunately, in the case relevant here,
there is no such supression of the small-$x_F$ region.
The percentage of energy not accounted for in the last column of
table~\ref{tab:gamma=1}
is a rough measure for the uncertainty resulting from this.
We see that~$\approx 49$\% of the power
is emitted as neutrinos where~$\approx 33$\% are
muon type neutrinos. The rest is electromagnetic where twice as much
luminosity is emitted in photons as in electrons.
Muon neutrinos and anti-neutrinos, which are the species which
experiments are most sensitive to, carry about~$2/3$ of the
electromagnetic power.

\subsection{Electromagnetic radiation}
\label{sec:em-reprocessing}

The electromagnetic output of the hadronic cascade is reprocessed in
the inner disk. Pair cascades, inverse Compton scattering, and
reflection on cold material change the shape of the initial
$E^{-2}$-spectrum and provide a steep turnover around
$E_\gamma \approx m_e \approx 511\keV$%
{}~\cite{zc:apj376,zggsfd:apjl363,ghi:mnras224,sve:mnras227,%
zdz:nasa-91-050,dgf:mnras245}.
This produces the observed X-ray and gamma emission of the AGN which
we will calculate in detail elsewhere.

\section{Resulting neutrino spectrum}

The neutrino spectrum from a single source mirrors the
$E^{-2}$-spectrum of the protons, only that it is shifted down by a
factor of $\epsilon_\nu \approx 0.05$. The upper cutoff of the
neutrino spectrum depends on the details of the shock acceleration
process in the jet, since that determines the maximum energy reached by
the protons~\cite{bie:hawaii1992}.

\paragraph{The Neutrino luminosity:}
To get an estimate of the neutrino luminosity of a source, we use its
total emission in X-rays and $\gamma$-rays as a reference.
We assume that the power emitted at X-ray and
$\gamma$-ray energies is the electromagnetic power output of the
hadronic cascade,
reprocessed in the inner disk to yield the observed spectrum
(see section~\ref{sec:em-reprocessing}).
The relation we get is then
\begin{equation}
  \dot \Etot_\nu = 0.95\, \dot \Etot_{\mathrm{X}+\gamma}.
\label{eqn:normalisation}
\end{equation}
Using an observation for $\Etot_{\mathrm{X}+\gamma}$,
equation~(\ref{eqn:sec-power}) allows us to determine~$C_\nu$. There is
actually a logarithmic dependence on the energy range of the neutrinos; for
definiteness we take the range\nolinebreak[3] $10^{-2}\GeV\dots 10^6\GeV$.

\subsection{X-ray background}

To be able to use the relation between electromagnetic output and
neutrino production to predict the diffuse neutrino background, we
have to estimate how much of the observed, diffuse X-ray and $\gamma$-ray
background results from hadronic cascades in AGN\@.
There is growing evidence from the ROSAT all sky survey that an appreciable
fraction of the background is due to active galactic nuclei, possibly a
dominant proportion~\cite{hst:aa246}.

The active nucleus in a quasar is often surrounded by a zone of very active
star formation.
ROSAT X-ray observations of
the Seyfert  galaxy NGC1068~\cite{welb:apj391} demonstrate that even
the~$2$--$10\keV$
X-ray  emission might be dominated by starburst activity near to the
nucleus.  On the other hand, X-ray observation of the starburst
galaxy M82~\cite{spbks:apj336}
show that the X-ray spectrum is rather hard and is likely
to contain an inverse Compton component, \ie, a powerlaw component.
Hence the starburst surrounding the active nucleus is likely to contribute a
hard powerlaw X-ray emission component.
What fraction of the total emission from a radioweak
quasar might then arise from this hard X-ray emission component?  In starburst
galaxies the radio ($5 \unit{GHz}$),
mid-infrared ($60 \unit{\mu m}$), and X-ray ($2 \keV$)
emission is correlated (see~\cite{ckb:aa219} for a comparison of
starburst galaxies with quasars), and thus the weakness of the radioemission
attributable to the starburst speaks against a dominant contribution from the
starburst region, as does the often observed variability of the X-rays emitted
by radioweak quasars.  On the other hand, this argument is not sufficiently
general to eliminate a strong contribution by a starburst in all quasars.

Thus we cannot decide conclusively to what extent a starburst
surrounding the active
nucleus in many radioweak quasars contributes hard X-ray emission
similar to that of the nuclear emission.
As a matter of fact, extensive work on compton reflection models for
AGN X-ray emission cannot explain all features of the observed X-ray
background spectrum~\cite{zzsb:crcxrb}.
To be conservative, we allow for a factor of three maximum
between the total hard X-ray emission from an
AGN galaxy (starburst and active nucleus together) and the contribution
strictly from the hadronic cascade.

The observations of the hard X-ray background at those photon energies
minimally influenced by reprocessing~\cite{fgmr:mnras242},
\ie, at energies where the original $E^{-2}$~powerlaw is still visible,
give a
possible range for the unreprocessed energy total of~$1.0 \cdot 10^5$
to~$1.4 \cdot 10^5 \eV \cm^{-2} \unit{s}^{-1} \unit{sr}^{-1}$.
Using the abovementioned conservative estimate, we arrive at a
lower limit of~$3 \cdot 10^4  \eV \cm^{-2} \unit{s}^{-1}\unit{sr}^{-1}$
for the contribution of hadronic cascades
to the X-ray and $\gamma$-ray background.

\subsection{Neutrino background}

Since the cosmological redshift is the same for
neutrinos and photons, we can scale the background neutrino luminosity
using the fraction of the X-ray and $\gamma$-ray background
derived above. Using equations%
{}~(\ref{eqn:sec-power}) and~(\ref{eqn:normalisation}), we get
\begin{equation}
  N(E_\nu) = 1.6 \cdot 10^{-12} \left(E_\nu \over \TeV\right)^{-2}
               \cm^{-2} \unit{s}^{-1} \unit{sr}^{-1} \GeV^{-1}
\end{equation}
as a conservative limit
for the sum of all neutrino species. About~$2/3$ are muon neutrinos,
the remaining~$1/3$ are electron neutrinos.
This prediction is a factor of $2.5$ lower than the experimental limit
set by the Frejus experiment~\cite{mey:frejus-limit,mey:private1}.
For each family, the
number splits evenly into neutrinos and anti-neutrinos.
Again, we have a weak dependence of the scale factor on the range of the
neutrino energies.
This expression is valid up to energies
of~$\approx 0.05 \Emax_p = 10^{7\pm1} \GeV$;
the spectrum turns over at this energy due to
the lack of parent particles to produce neutrinos.
Actually, we
expect the background spectrum to have a less sharp cutoff at high
energies than a single source,
since the maximum neutrino energy varies between different
AGN\@. Therefore, the cutoff in the superposition of all spectra
will be smoothed out.

Such a spectrum is similar to the results of the Monte-Carlo calculation
in~\cite{sp:hawaii1992}; but the flux of muon
neutrinos predicted here is about one order of magnitude lower.
Compared to the prediction for the neutrino
background made in~\cite{sdss:hawaii1992}, we expect the
$E^{-2}$~powerlaw for the neutrino spectrum down to the
$\GeV$~range without the flattening seen in~\cite{sdss:hawaii1992}
below~$10^6\GeV$ --- a difference which should be visible in the
experiment.

\section{Discussion}

The main point in our model is the production of neutrinos as the
result of a hadronic cascade. Compared to the
$p\gamma$-channel, which has a threshold of
\mbox{$E_p\approx 8\cdot10^6\GeV$} for photoproduction
on UV~photons,
all protons above a few hundred~$\MeV$ contribute in $pp$-interactions. As a
consequence, we expect neutrino emission even from sources with a low
cutoff in the primary proton spectrum.
This way, we include contributions from a larger class of sources both
to the neutrino background and to the diffuse X-ray background.
Similar to the result of other authors~\cite{sdss:hawaii1992,sp:hawaii1992},
our model displays the tendency to produce neutrinos strikingly close
to existing limits~\cite{mey:frejus-limit,mey:private1,hz:pl289b}.

Modifications of this $pp$-model can lead to a reduced neutrino flux
while reprocessing of the electromagnetic component ensures an
unchanged X-ray emission. The most drastic modification --- reducing
the maximum proton energy and correspondingly the maximum neutrino
energy --- can ultimately decrease the observable extragalactic
neutrino flux by moving the cutoff below the cross-over with the
steeper spectrum of atmospheric neutrinos. More realistically,
steepening of the AGN proton distribution at a sufficiently low break
energy is an alternative to reduce the predicted event rate in a
neutrino detector. In both cases, neutrino production via
$p\gamma$-reactions becomes ineffective.

On the other hand, a proton spectrum flatter than~$E^{-2}$ leads to
the dominance of $p\gamma$-reactions at high energies. This, again
keeping the X-ray background unchanged, enhances the neutrino flux far
above the $\TeV$ range while reducing the flux below. Details depend
on the modelling of the target photon field (\ie, contributions from
the disk, corona, jet, \etc).

Another important point in our model is that the acceleration of the
protons takes place above the disk.
This way, the acceleration takes place in an environment more suitable
than an accretion disk.

\paragraph{Acknowledgements:}
We would like to thank V.~S.~Berezinsky,
F.~Halzen, T.~K.~Gaisser, H.~Meyer,
M.~Niemeyer, T.~Stanev,
and F.~W.~Stecker for helpful discussions.


\newcommand{\noopsort}[1]{} \newcommand{\switchargs}[2]{#2#1}

\end{document}